\documentclass{aastex631}
\usepackage{epstopdf}
\usepackage{natbib}
\usepackage{graphicx}
\usepackage{CJK}

\usepackage[percent]{overpic}
\usepackage{times}
\usepackage{graphicx}
\usepackage{amsmath}
\usepackage{subfigure}
\usepackage{natbib}
\usepackage{txfonts}
\usepackage{journals}
\usepackage{xcolor}
\usepackage[utf8]{inputenc}
\usepackage[overload]{empheq}
\usepackage{verbatim}

\newcommand{\ks}{$K_{\rm S}$}

\newcommand{\gbp}{$G_{\rm BP}$}
\newcommand{\grp}{$G_{\rm RP}$}

\newcommand{\ebr}{$E(G_{\rm BP}-G_{\rm RP})$}

\newcommand{\ebv}{$E(B-V)$}

\newcommand{\mg}{$M_{G}$}

\newcommand{\teff}{$T_{\rm eff}$}
\newcommand{\feh}{$\rm [Fe/H]$}
\newcommand{\logg}{${\rm log}\,g$}

\submitjournal{AJ}
\begin{document}
\begin{CJK*}{UTF8}{gbsn}

\title[Stellar parameters from multi-band photometries]
{SPar: estimating stellar parameters from multi-band photometries with empirical stellar libraries}

\author[0000-0002-2473-9948]{Mingxu Sun (孙明旭)}
\affiliation{Department of Physics,
               Hebei Key Laboratory of Photophysics Research and Application, 
               Hebei Normal University,
               Shijiazhuang 050024, P.\,R.\,China}
               
\author[0000-0003-2472-4903]{Bingqiu Chen(陈丙秋)}
\affiliation{South-Western Institute for Astronomy Research,
               Yunnan University,
               Kunming 650500, P.\,R.\,China}

\author[0000-0001-5737-6445]{Helong Guo(郭贺龙)}
\affiliation{South-Western Institute for Astronomy Research,
               Yunnan University,
               Kunming 650500, P.\,R.\,China}   

\author[/0000-0003-2645-6869]{He Zhao(赵赫)}
\affiliation{Purple Mountain Observatory,
               Chinese Academy of Sciences,
               Nanjing 210023, P.\,R.\,China}

\author[0000-0001-8247-4936]{Ming Yang(杨明)}
\affiliation{Key Laboratory of Space Astronomy and Technology,
               National Astronomical Observatories,
               Chinese Academy of Sciences,
               Beijing 100101, P.\,R.\,China}
               
\author[0000-0003-1359-9908]{Wenyuan Cui(崔文元)}
\affiliation{Department of Physics,
               Hebei Key Laboratory of Photophysics Research and Application, 
               Hebei Normal University,
               Shijiazhuang 050024, P.\,R.\,China}        
\correspondingauthor{Bingqiu Chen}
\email{bchen@ynu.edu.cn}

\date{Accepted 08-Jul-2023. Received 18-Jun-2023; in original form 23-May-2023}

\label{firstpage}

\begin{abstract}
Modern large-scale photometric surveys have provided us with multi-band photometries of billions of stars. Determining the stellar atmospheric parameters, such as the effective temperature (\teff) and metallicities (\feh), absolute magnitudes ($M_{G}$), distances ($d$) and reddening values (\ebr) is fundamental to study the stellar populations, structure, kinematics and chemistry of the Galaxy. This work constructed an empirical stellar library which maps the stellar parameters to multi-band photometries from a dataset with Gaia parallaxes, LAMOST atmospheric parameters, and optical to near-infrared photometry from several photometric surveys. Based on the stellar library, we developed a new algorithm, SPar (\textbf{S}tellar \textbf{P}arameters from multib\textbf{a}nd photomet\textbf{r}y), which fits the multi-band stellar photometries to derive the stellar parameters (\teff, \feh, $M_G$, $d$ and \ebr) of the individual stars. The algorithm is applied to the multi-band photometric measurements of a sample of stars selected from the SMSS survey, which have stellar parameters derived from the spectroscopic surveys. The stellar parameters derived from multi-band photometries by our algorithm are in good agreement with those from the spectroscopic surveys. The typical differences between our results and the literature values are 170\,K for \teff, 0.23\,dex for \feh, 0.13\,mag for $M_G$ and 0.05\,mag for \ebr. The algorithm proved to be robust and effective and will be applied to the data of future large-scale photometric surveys such as the Mephisto and CSST surveys.
\end{abstract}

\keywords{dust, extinction -- stars: low-mass -- solar neighbourhood}

\section{Introduction} \label{sec:intro}

Modern large-scale photometric surveys, such as the Sloan Digital Sky Survey (SDSS; \citealt{SDSS2000}), the Pan-STARRS 1 Survey (PS1; \citealt{Chambers2016}),  the Two Micron All Sky Survey (2MASS; \citealt{Skrutskie2006}) and the Wide-Field Infrared Survey Explorer (WISE; \citealt{WISE2010}), have provided us board band photometry  covering the wavelength from the optical to the infrared (IR) bands of billions of stars. The multi-band photometric measurements of stars contain the spectral energy distribution (SED) information of these stars, from which we can obtain the stellar atmosphere parameters (i.e. the effective temperature \teff, the metallicity \feh\ and the surface gravity log\,$g$), as well as the distance and extinction values of stars \citep{Berry2012, Chen2014, Green2014, Green2015}.

Stellar atmospheric parameters are typically determined from their spectra. While large-scale multi-fibre spectroscopic surveys like the Large Sky Area Multi-Object Fiber Spectroscopic Telescope (LAMOST; \citealt{Luo2015}), the Apache Point Observatory Galactic Evolution Experiment (APOGEE; \citealt{SDSSDR17}), the Galactic Archaeology with HERMES (GALAH, \citealt{Buder2021}), and the Dark Energy Spectroscopic Instrument (DESI; \citealt{2022AJ....164..207D}) have provided spectra for tens of millions of stars, we can now determine the parameters of billions of stars with comparable precision using photometric data of exceptionally high accuracy or data obtained through specially designed filters. This approach allows us to obtain accurate stellar parameters without relying solely on spectroscopic data. For example, \citet{Xu2022} have measured metallicities of $\sim$ 27 million stars from the Gaia Early Data Release 3 (Gaia EDR3; \citealt{Gaiaedr3-2021}) photometric data of unprecedented millimagnitude precision. The typical metallicity precision is about $\delta$[Fe/H] = 0.2\,dex. Based on the narrowband photometry from the SkyMapper Southern Survey (SMSS; \citealt{Wolf2018, Onken2019}), \citet{Huang2022} have determined stellar atmospheric parameters for $\sim$ 24 million stars. The precision of their metallicity estimates has typical values around 0.05 to 0.15\,dex. \citet{Yang2022} obtained stellar parameters (effective temperature, \teff, surface gravity, \logg, and metallicity, \feh) from the narrow-band photometries of the J-PLUS survey \citep{Cenarro2019}. They have achieved precisions of $\delta$\teff $\sim$ 55\,K, $\delta$\logg $\sim$ 0.15\,dex, and $\delta$\feh $\sim$ 0.07\,dex, respectively. We note that these works are only for stars located at high Galactic latitudes, where the interstellar extinction is small. For stars at low Galactic latitudes, where the extinction effects are large, the stellar atmospheric parameters need to be estimated along with the stellar extinction values.

The future Multi-channel Photometric Survey Telescope (Mephisto; \citealt{Yuan2020}) photometric survey and the Chinese Space Station Telescope (CSST; \citealt{Zhan2011, Zhan2018}; \citealt{Cao2018}) optical survey will have both high precision and specially designed filters, which will provide great opportunities for us to obtain accurate stellar parameters of  billions of stars. Mephisto is a wide-field survey telescope with a 1.6\,m primary mirror. It is equipped with three CCD cameras and is capable to image the same patch of sky in three bands simultaneously, which will provide us with real-time colours of stars with unprecedented accuracy. The filters of Mephisto ($uvgriz$) are similar to those of SkyMapper \citep{Bessell2001}. The Mephisto-W Survey (\citealt{Lei2021, Lei2022}; Chen et al. in prep.) will target the northern sky of declination (Dec) between $-$21\degr\ and 75\degr\ with a coverage of over 27,000 deg$^2$. CSST is a 2\,m space survey telescope, which shares the same orbit as the Chinese Space Station. The CSST optical survey (CSST-OS) will observe a large sky area of $\sim$ 18,000\,deg$^2$ in seven photometric filters ($NUV, ~u, ~g, ~r, ~i, ~z$ and $y$) covering the wavelengths from the near-ultraviolet (NUV) to near-infrared (NIR).

Deriving the stellar atmospheric parameters as well as the distance and extinction values is a fundamental task for the Mephisto and CSST surveys. In this work, we present a new algorithm，SPar (\textbf{S}tellar \textbf{P}arameters from multib\textbf{a}nd photomet\textbf{r}y)， to estimate stellar atmospheric parameters, such as the effective temperature (\teff) and metallicities (\feh), absolute magnitudes ($M_{G}$), distances ($d$) and reddening values (\ebr) of a large sample of stars from their multi-band photometries. Previous algorithms such as the Star-Horse \citep{Anders2019, Anders2020} and the General stellar Parameterized from photometry (GSP-PHot; \citealt{Andrae2022}) rely on the theoretical stellar models, which may suffer systematic effects \citep{Green2021}. One method of dealing with these inaccuracies in theoretical models is to apply empirical corrections based on the observed photometry of stars of known type. \citet{Berry2012}, \citet{Chen2014} and \citet{Green2015} measure stellar parameters and extinction based on the empirical stellar locus in the colour-colour space. \citet{Sun2018}, \citet{ Chen2019} and \citet{Sun2021} calculated the intrinsic colours and reddening values of the individual stars based on a spectroscopic sample selected from the spectroscopic surveys. In this work, we will construct an empirical stellar library which maps the stellar parameters to multi-band photometries from a dataset with Gaia parallaxes, LAMOST atmospheric parameters, and optical to near-infrared photometry from several photometric surveys. 

The paper is structured as follows: Section~\ref{sec:dat} presents the relevant dataset. Section~\ref{sec:emp} describes in detail the empirical stellar library we constructed. Section~\ref{sec:est} describes our algorithm and Section~\ref{sec:tes} tests it. Finally, Section~\ref{sec:sum} summarizes the algorithm.

\section{Data}\label{sec:dat}

As the Mephisto and CSST surveys have not yet started, in the current work we have used photometric data from the SMSS survey for the experiment. This work is based on the Gaia Data Release 3, the broad-band photometry from SMSS, Two Micron All Sky Survey (2MASS; \citealt{Skrutskie2006}) and the Wide-field Infrared Survey Explorer survey (WISE; \citealt{WISE2010}), and the spectroscopic data from LAMOST, APOGEE and GALAH.

The SMSS is an ongoing photometric survey of the Southern sky \citep{Wolf2018, Onken2019}. The survey depth is between 19.7 and 21.7\,mag in six optical bands: $u,~v,~g,~r,~i$ and $z$. In this work, we adopt the data from its second data release (SMSS DR2; \citealt{Onken2019}). The photometry has a internal precision of 1\,per\,cent in the $u$ and $v$ bands, and 0.7\,per\,cent in the other four bands ($g,~ r,~ i$ and $z$).

To break the degeneracy of effective temperature (or intrinsic colours) and extinction for the individual stars, we combine the SMSS photometry with the IR photometry of 2MASS and
WISE. The 2MASS survey is a full-sky survey undertaken in three filters: $J, ~H$ and \ks. The systematic errors of the 2MASS photometry are estimated to be less than 0.03\,mag \citep{Skrutskie1997}. The WISE survey is a full-sky survey undertaken in four bands: $W1, ~W2, ~W3$ and $W4$. In the current work, we adopt the AllWISE Source Catalog \citep{Kirkpatrick2014}. We use only the data in the $W1$ and $W2$ bands, as the $W3$ and $W4$ measurements have lower sensitivities and poorer angular resolutions.

The Gaia DR3 \citep{Gaiadr3} photometric data and parallax measurements are also adopted to derive the stellar parameters when available. Gaia DR3 was released by the Gaia mission \citep{Gaiamission}.  It contains more than a billion sources with five astrometric parameters (position, parallaxes, and proper motions) and three-band photometry ($G$, \gbp\ and \grp).  The median uncertainty of the parallax is $\sim$ 0.02 - 0.03\,mas for G $<$ 15\,mag sources, 0.07\,mas at G = 17\,mag, and 0.5\,mas at G = 20 mag \citep{Gaiaedr3-2021}. At G = 20\,mag, the typical uncertainties for the Gaia DR3 photometries are 6, 108, and 52\,mmag for the Gaia $G$, \gbp\ and \grp\ bands, respectively \citep{Gaiaedr3-2021}. 

The LAMOST, APOGEE and GALAH spectroscopic data are adopted in the current work for two aims: to construct the empirical stellar library and to validate the resulting stellar properties. In this work, we use the `LAMOST LRS Stellar Parameter Catalog of A, F, G and K Stars' from the LAMOST data release 8 (LAMOST DR8; \citealt{Luo2015}). The catalogue contains stellar atmospheric parameters (\teff, \logg\ and \feh) derived from over 6 million low-resolution spectra. For the APOGEE data, we use the APOGEE stellar parameters catalogue from the SDSS data release 17 (SDSS DR17; \citealt{SDSSDR17}). The catalogue contains stellar atmospheric parameters (\teff, \logg\ and [M/H]) derived from over 0.6 million near-IR spectra. For the GALAH data, we adopt its DR3 catalogue \citep{Buder2021}, which contains stellar parameters (\teff, \logg\ and [Fe/H]) of over 0.5 million nearby stars.

\section{Empirical stellar library}\label{sec:emp}

An empirical stellar library is first created based on a sample of stars selected from the LAMOST spectroscopic data and Gaia DR3, for which atmospheric parameters, distances, and extinction values can be well measured. We cross-match the LAMOST and Gaia stars with the optical and near-IR photometric data, i.e. the SMSS, 2MASS and WISE, using a radius of 1\,arcsec. To exclude the stars with bad observations, a sample of stars are selected by the criteria as below:

 \begin{enumerate}
 \item LAMOST spectral signal-to-noise ratio (SNR) larger than 20 and effective temperature \teff\ between 4000 and 8000\,K,
 \item All Gaia $G$\gbp\grp\  photometric errors smaller than 0.1\,mag and 
 phot\_bp\_rp\_excess\_factor $>$ 1.3 + 0.06(\gbp\ - \grp)$^2$,
 \item Photometric errors in any SMSS $uvgriz$, 2MASS $JH$\ks\ and WISE $W1W2$ bands less than 0.1\,mag.
 \end{enumerate}
 
To obtain the absolute magnitude of our selected stars in each filter, the extinction values in the individual filters and the distance of each star need to be calculated.

\subsection{Correct the extinction effect of the individual stars}
In this work, The reddening values of our sample stars are calculated with a star-pair method \citep{Yuan2015}. In selecting the control sample for extinction correction, this study impose more rigorous constraints on the signal-to-noise ratio (SNR) of the LAMOST spectrum and photometric accuracies compared to the empirical stellar library, and only low-extinction stars are chosen. The control stars are selected via the following criteria: 

 \begin{enumerate}
 \item LAMOST spectral signal-to-noise ratio (SNR) larger than 50,
 \item Gaia $G$\gbp\grp\  photometric errors smaller than 0.01\,mag, all SMSS $uvgriz$, 2MASS $JH$\ks\ and WISE $W1W2$ bands photometric errors less than 0.08\,mag,
 \item \ebv\ values from the extinction map of \citet{Planck2014} smaller than 0.025\,mag.
 \end{enumerate}
 
For a given target star in our sample, its intrinsic colours, $(G_{\rm BP} - x)_0$ (where $x$ denotes to the magnitude in another band, i.e., $G$, \grp\,$u,~v,~g,~r,~i,~z,~J,~H$, \ks, $W1$ or $W2$), are estimated simultaneously from the corresponding values of the pair stars in the control sample. The reddening values of $E(G_{\rm BP} -x)$ of the target star are then obtained from the differences between its observed and intrinsic colours, i.e. $E(G_{\rm BP} -x) = (G_{\rm BP} - x) - (G_{\rm BP} - x)_0$.

Stars with resulted $E(G_{\rm BP} -x)$ errors larger than 0.1\,mag are excluded from our catalogue. Based on the resultant reddening values $E(G_{\rm BP} -x)$, a \teff\ and reddening dependent extinction law similar to the work of \citet{Zhang2023} has been built. \citet{Zhang2023} obtained the empirical reddening coefficients for the individual colours as a function of \teff\ and \ebv. In the current work, we derive the empirical reddening coefficients, defined as $R(G_{\rm BP} - x) = \dfrac{E(G_{\rm BP} - x)}{E(G_{\rm BP} - G_{\rm RP})}$, as a function of \teff\ and \ebr.  In the current work, we adopt a binary function: 

\begin{equation}\label{eq:r}
R(G_{\rm BP} - x) = C_{0}+C_{1}x+C_{2}x^2+C_{3}y+C_{4}xy+C_{5}y^2, 
\end{equation}
where $x = T_{\rm eff} - 6000$ and $y = E(G_{\rm BP}-G_{\rm RP}) - 0.5$. The resulting reddening values we derived from the star-pair method are used to fit Eq.~\ref{eq:r} to obtain the individual coefficients $C_0$ to $C_5$. The fitting results are listed in Table~\ref{tab1}.

\begin{figure*}
\centering
\includegraphics[height=7.8cm]{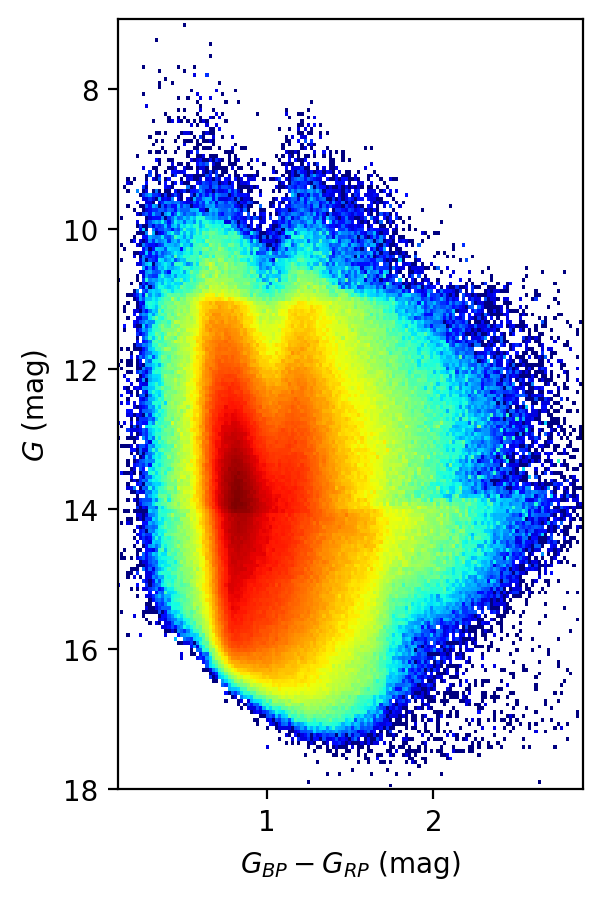}
\includegraphics[height=7.8cm]{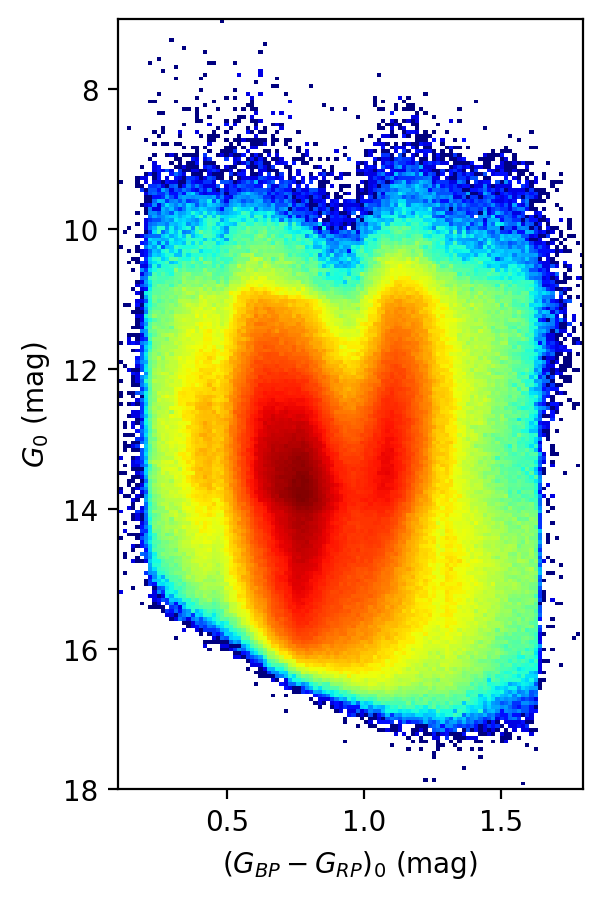}
\includegraphics[height=7.8cm]{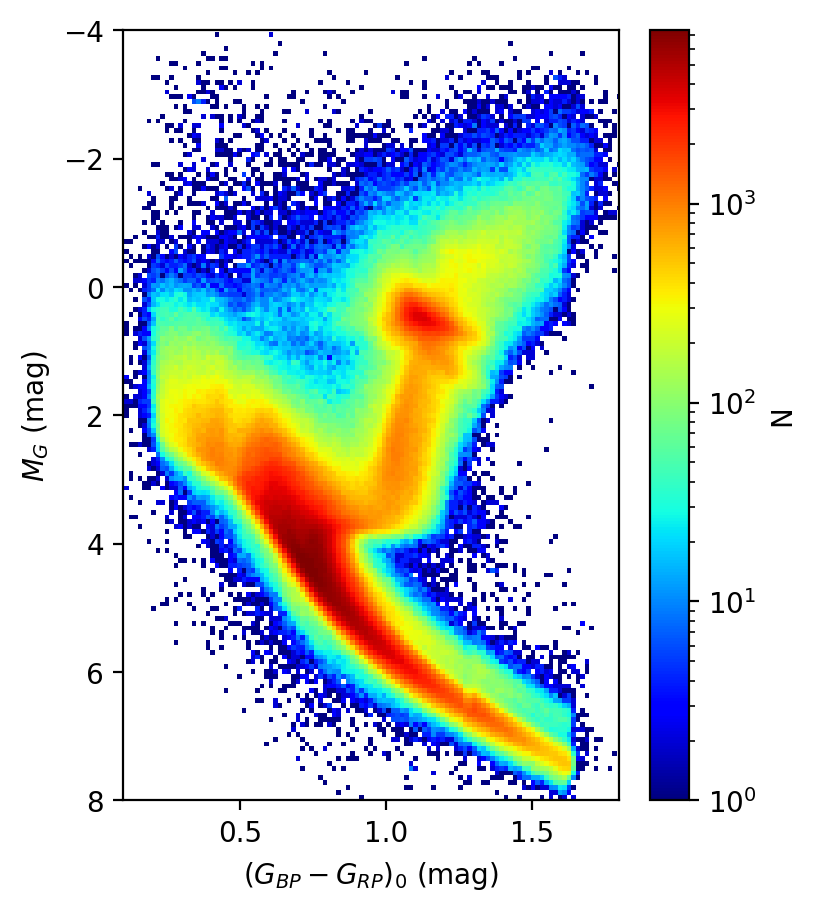}
\caption{{\it Left panel:} Gaia observed colour - magnitude diagram. {\it Middle panel:} Gaia de-reddened colour-magnitude diagram. {\it Right panel}: Gaia HR diagram.}
\label{cmdc}
\end{figure*}

 Finally, we assume that  $A_{W2}/E(G_{\rm BP}-G_{\rm RP}) = 0.063$ \citep{Wang2019} for the $W2$ band has the longest wavelength and experiences the least amount of extinction of all the bands.
 Combining  with the extinction coefficient relations we derived, the extinction value in each filter for all our sample stars can be calculated from their \ebr\ reddening values. The extinction values are then subtracted to obtain the intrinsic magnitude of the stars. In the left and middle panels of Fig.~\ref{cmdc}, we show the observed and intrinsic colour and magnitude diagrams (CMD) of our sample stars, respectively.

\subsection{The empirical HR diagrams}

We then obtain the absolute magnitudes of the sample stars. The distances of the sample stars are calculated from the Gaia DR3 parallaxes via a simple Bayesian approach \citep{BailerJones2018, Chen2019OBStar, Shen2022}. A simple posterior probability is adopted:

 \begin{equation}
    p(d|\varpi) = d^2\exp(-\dfrac{1}{2\sigma^2_{\varpi}}(\varpi-\varpi_{\rm zp} - \dfrac{1}{d}))p(d),
\end{equation}
where $\sigma_{\varpi}$ and $\varpi _ {\rm zp}$ are respectively the errors and globular zero points of the Gaia parallaxes, and $p(d)$ the space density distribution prior for the sample stars. In the current work, a zero point of $\varpi _ {\rm zp}$ $=$ $-$0.026\,mas from \citet{Huang2021} is adopted. The Galactic structure model of \citet{Chen2017} is adopted as the spatial density distribution prior. With the resultant distances, the absolute magnitude in each band is then calculated by $M_x = x - 5\log\,d + 5 - A_x$. Stars with parallax errors larger than 20\,per\,cent are excluded. This leads to a final sample of 3,842,671 stars. In the sample, more than 3.5 million stars have all Gaia absolute magnitude estimates, over 3.2 million stars have all IR bands (2MASS $JH$\ks\ and WISE $W1W2$) absolute magnitudes, and over 0.3 million stars have all SMSS absolute magnitude estimates. In the right panel of Fig.~\ref{cmdc}, we show the resulting Hertzsprung–Russell diagram (HRD) of our final sample stars. 

\begin{figure*}
\centering
\includegraphics[height=7.8cm]{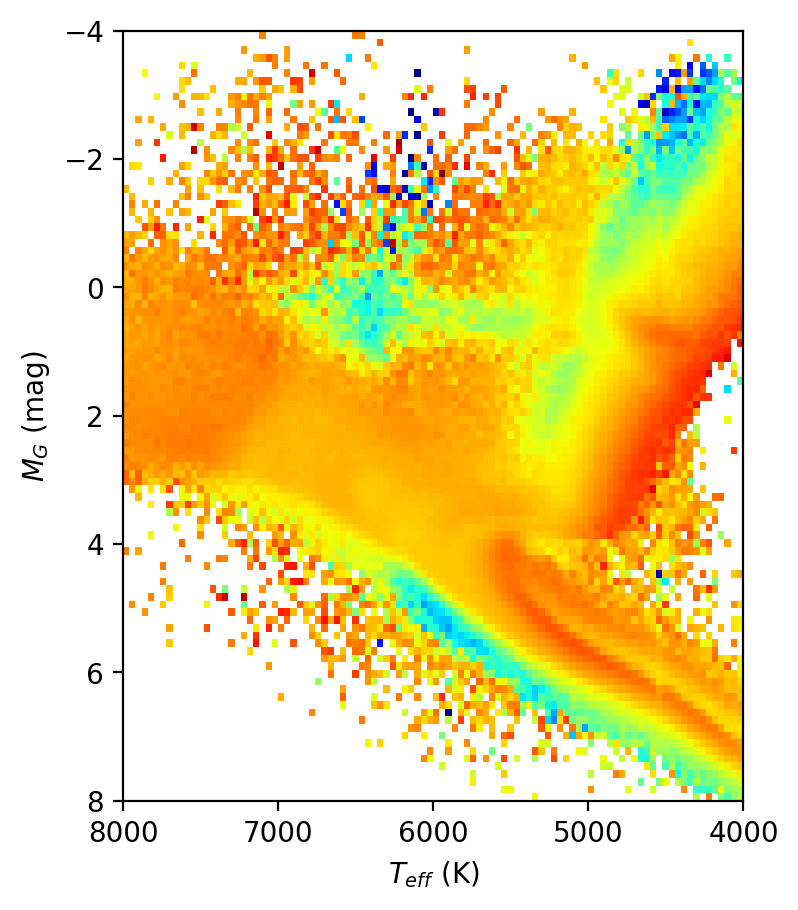}
\includegraphics[height=7.8cm]{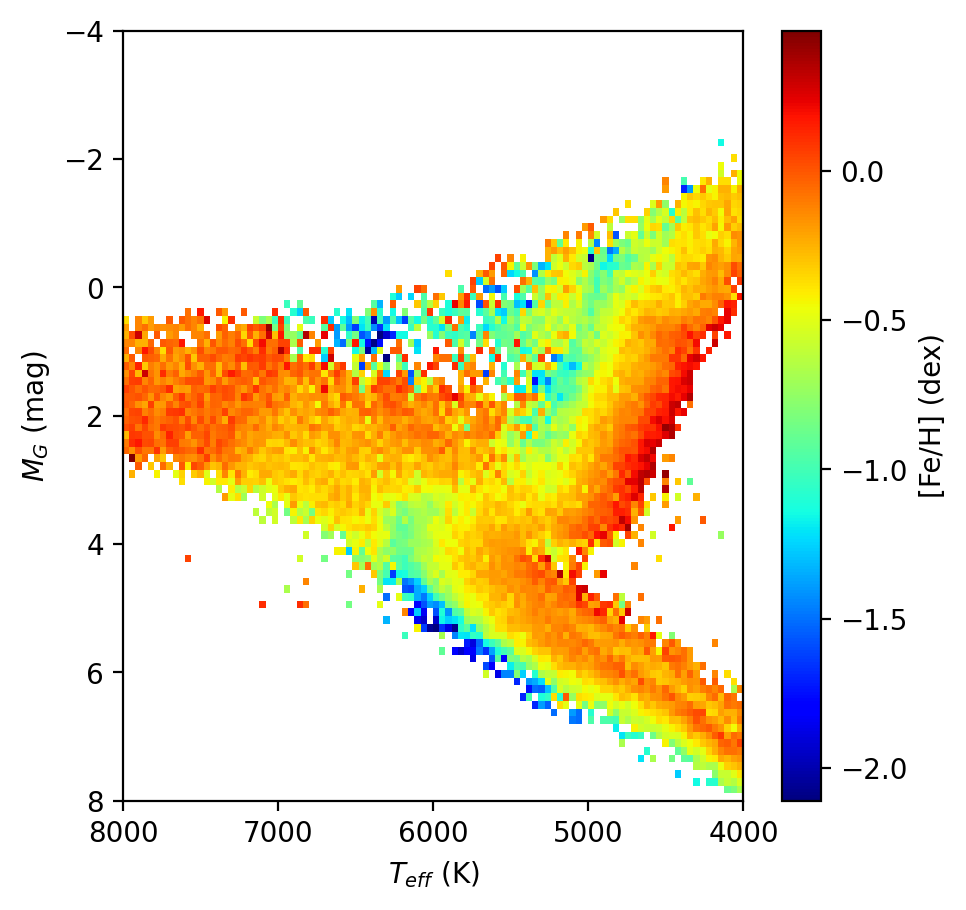}
\caption{The HRDs plotted with our final sample stars (left) and the resultant gridded stellar library (right). The colour scales represent the mean metallicities of each \teff\ - $M_G$ bin.}
\label{hrds}
\end{figure*}

The final sample of stars we obtained above is too large and not evenly distributed across the parameter spaces. Therefore, we use this sample to create a gridded stellar library. To map the stellar parameters, namely effective temperature (\teff), metallicity (\feh), and Gaia $G$ band absolute magnitude ($M_G$), to absolute magnitude in each filter, we divide the parameter spaces into 100 bins for \teff\ (ranging from 4000 to 8000 K), 50 bins for \feh\ (ranging from $-$2.5 to 0.5 dex), and 100 bins for $M_G$ (ranging from 8 to $-$4 mag). We use a machine learning algorithm called Random Forest regression to obtain the absolute magnitudes in each passband for each bin, using the final sample stars as the training dataset. We exclude any grids with fewer than 5 stars, resulting in an empirical stellar library with 39,905 grids in the parameter space. In Fig.~\ref{hrds}, we present the Hertzsprung-Russell diagrams (HRDs) of both the final sample stars and the resultant gridded stellar library.

\begin{figure*}
\centering
\includegraphics[width=0.19\textwidth]{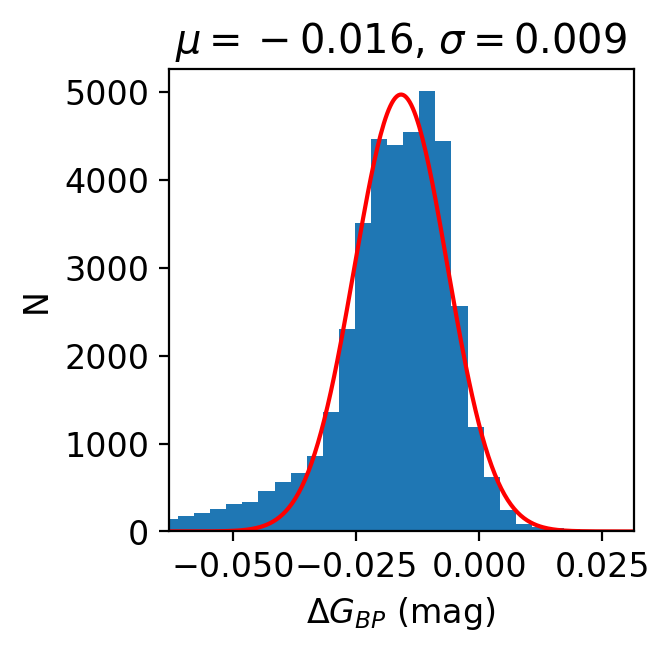}
\includegraphics[width=0.19\textwidth]{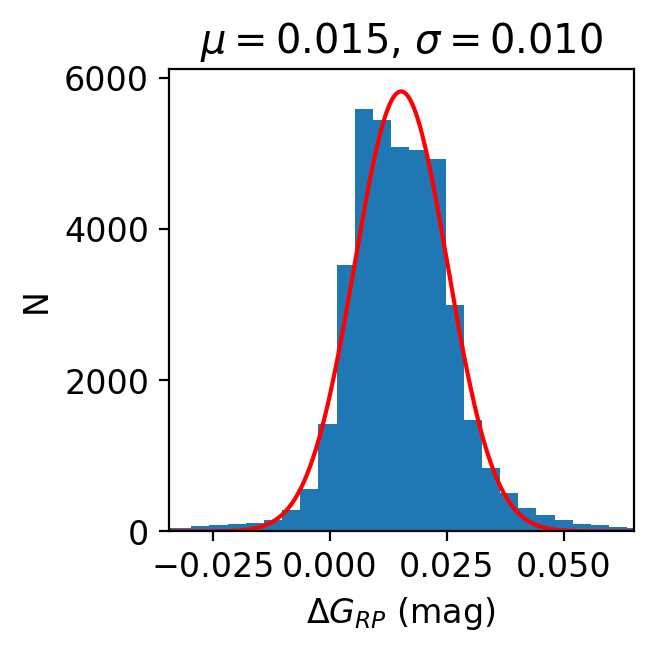}
\includegraphics[width=0.19\textwidth]{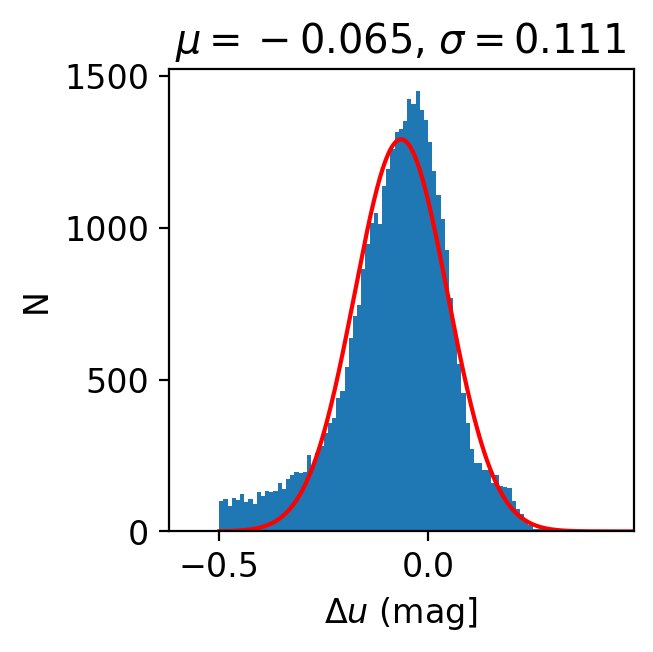}
\includegraphics[width=0.19\textwidth]{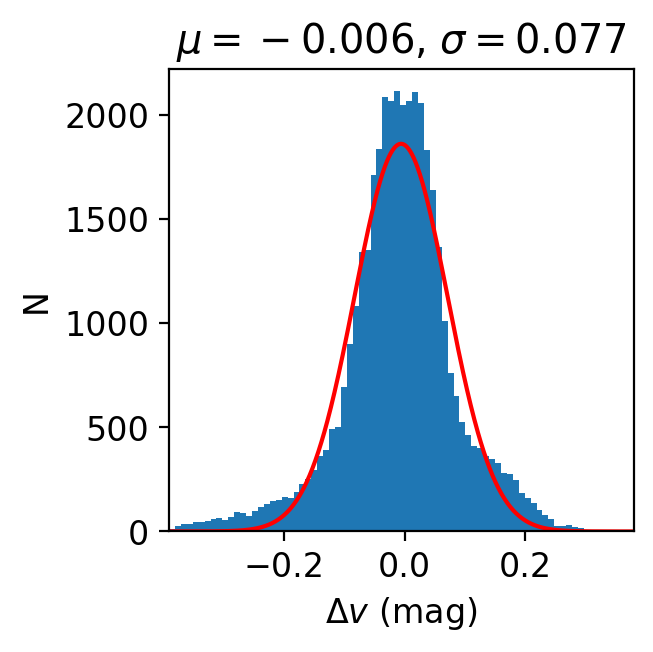}
\includegraphics[width=0.19\textwidth]{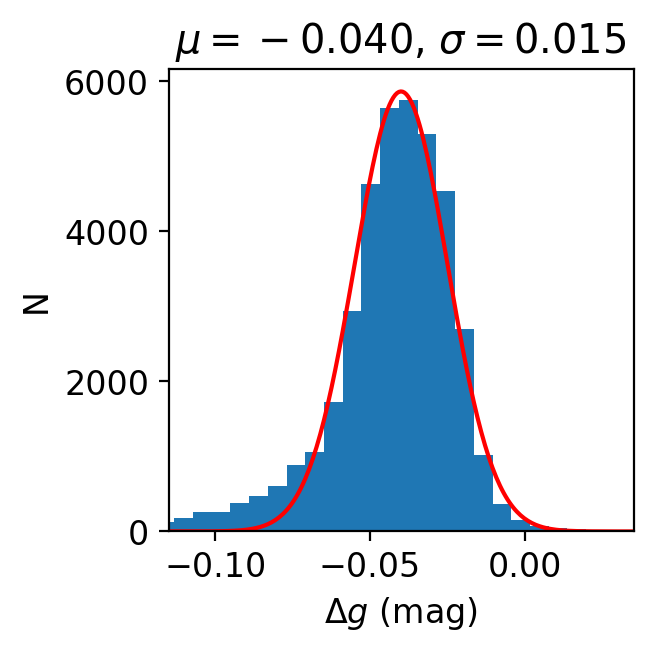}
\includegraphics[width=0.19\textwidth]{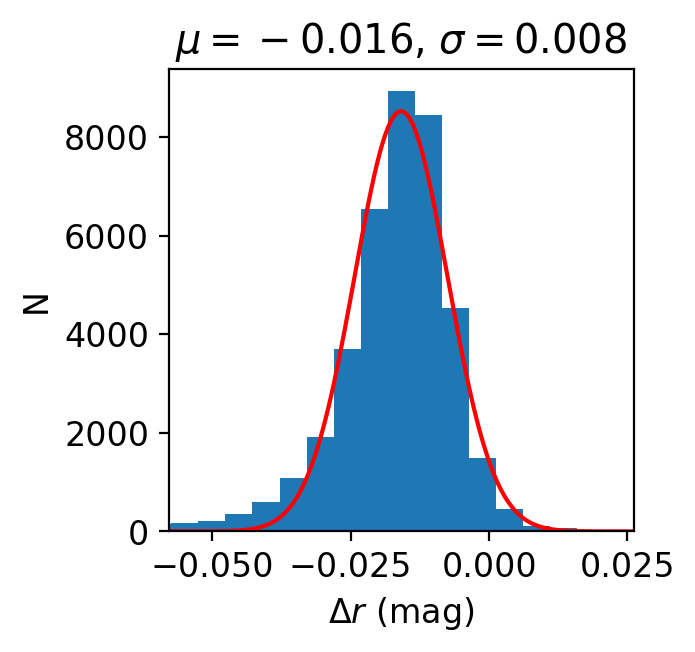}
\includegraphics[width=0.19\textwidth]{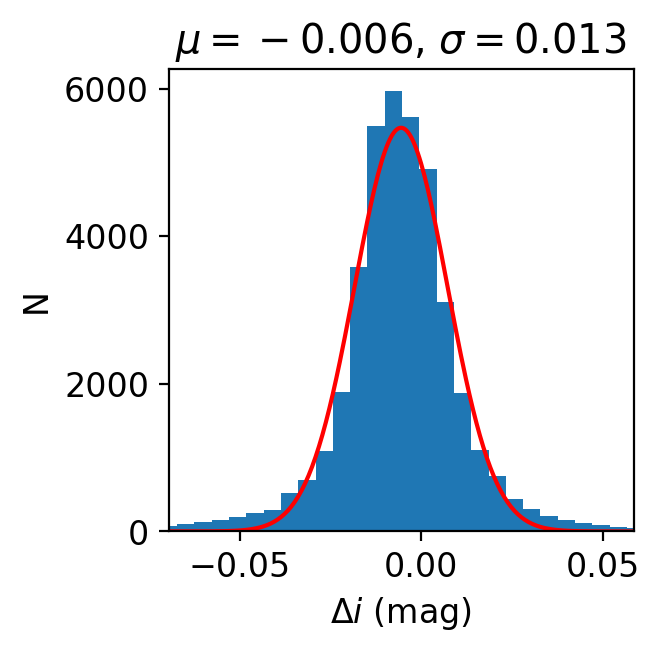}
\includegraphics[width=0.19\textwidth]{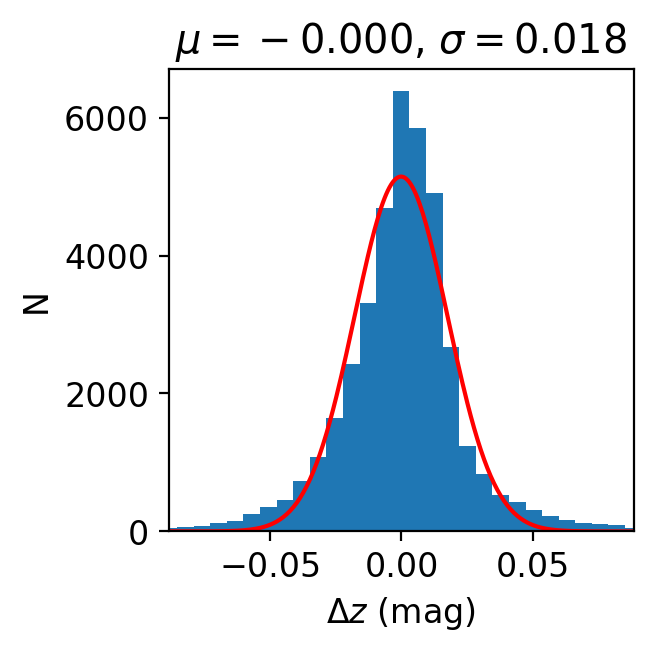}
\includegraphics[width=0.19\textwidth]{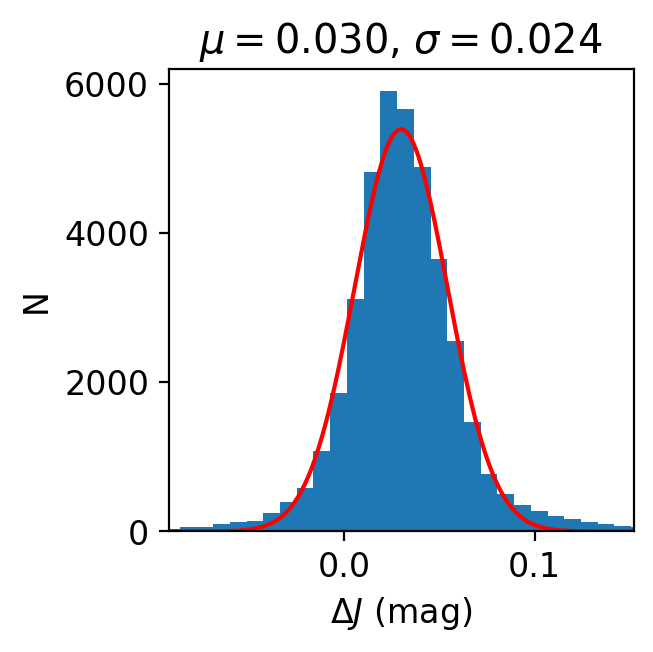}
\includegraphics[width=0.19\textwidth]{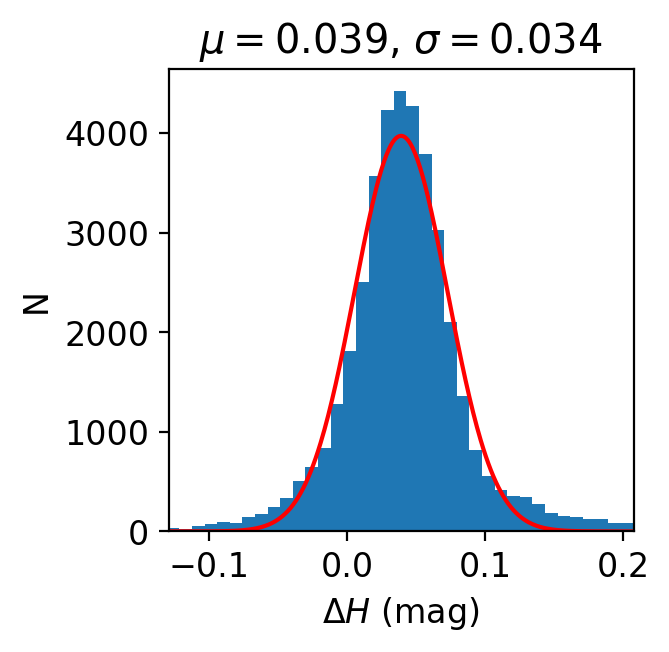}
\includegraphics[width=0.19\textwidth]{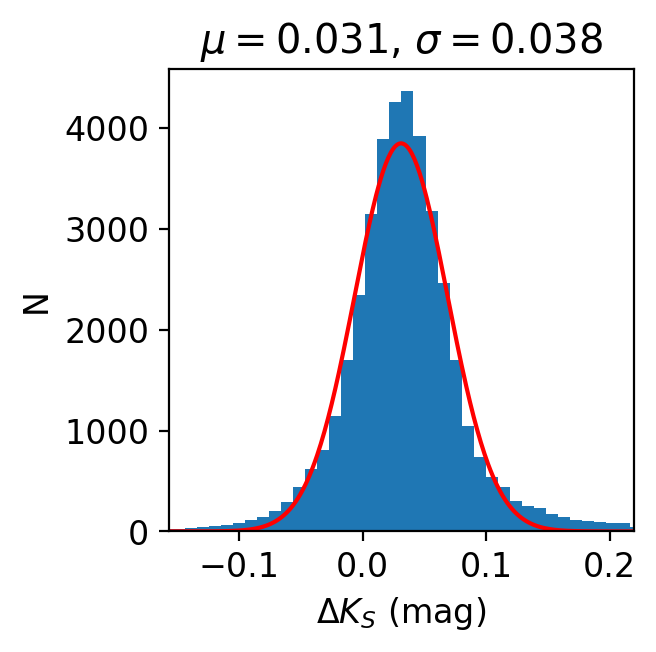}
\includegraphics[width=0.19\textwidth]{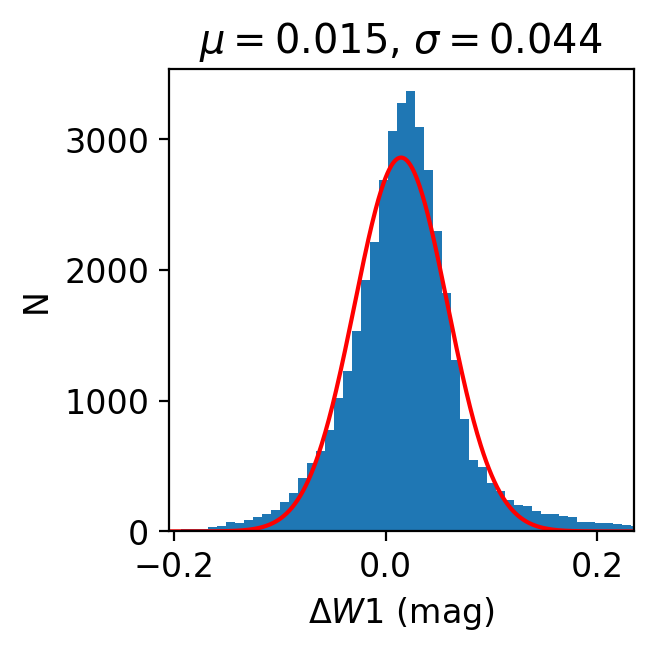}
\includegraphics[width=0.19\textwidth]{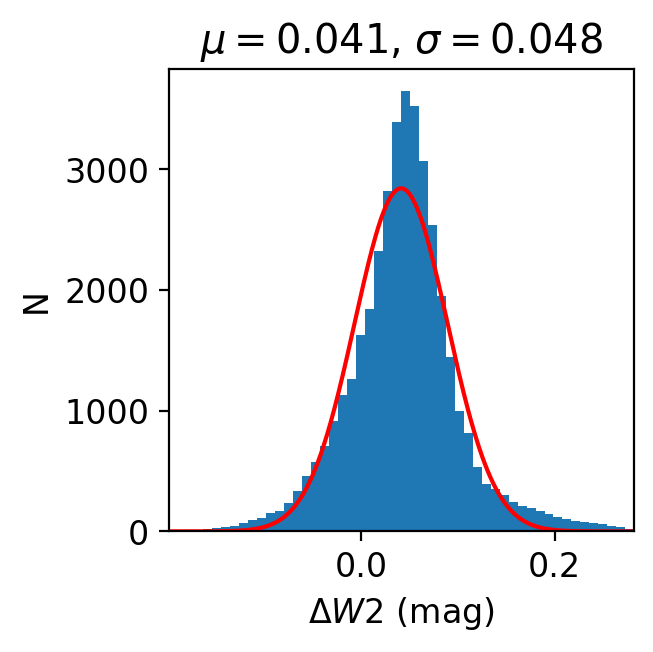}
\caption{Histograms of the differences between our predicted absolute magnitudes of stars and those from the PARSEC isochrones \citep{PAdova2012}. The mean and dispersion values of the differences are labelled in each panel.}
\label{isoc}
\end{figure*}

We performed a comparative analysis between our empirical stellar library and the PARSEC theoretical stellar isochrones \citep{PAdova2012} as a means of verifying our results. To achieve this, we linearly interpolated the PARSEC absolute magnitudes into our \teff, \feh\ and $M_G$ grids, and subsequently compared them with our corresponding results. As illustrated in Fig.~\ref{isoc}, the results of this comparison indicate that our stellar library is in good agreement with the PARSEC theoretical models. Specifically, the mean values of the differences between our empirical library and the PARSEC models ranged between $-$0.04 and 0.04\,mag, with dispersions of  0.01 to 0.03\,mag observed for most of the filters. We note a slight increase in the dispersions for the WISE $W1W2$ bands, which ranged between 0.04 - 0.05\,mag, and the SMSS $uv$ bands, which show dispersions of 0.08 - 0.11\,mag. We attribute the increase in dispersion to the relatively large calibration errors and uncertainties associated with the filter response curves of these particular filters. For the $u$ band, we find a slightly higher mean difference of $-$0.065 mag than other filters. This difference could be due to a variety of factors, including notable photometric error, extinction, and model uncertainty in the passband.

\section{Estimating stellar parameters from multi-band photometries}\label{sec:est}

This section introduces how we derive the stellar parameters from the multi-band photometries. To allow our algorithm SPar to be applied to large samples of billions of stars, we need to minimise the computational cost. Our algorithm fits only four stellar parameters, \teff, \feh, \mg\ and \ebr. The distances $d$ of stars can be further derived from the fitting results. SPar uses an ensemble Markov-chain Monto-Carlo (MCMC) method to obtain the best parameters of the individual stars by adopting a set of initial values derived from a minimum $\chi^2$ method.

\subsection{Initial parameters from the minimal $\chi^2$ method}\label{sec:chi}

Based on the reference stellar library and extinction law derived in Sect.~\ref{sec:emp}, assuming a reddening value, we can predict the `distance module corrected' magnitudes in the individual filters $M^\prime_x$ of stars, i.e., $M^\prime_x = M_x + A_x$. Distance module $\mu$ can then be derived by subtracting $M^\prime_x$ from the observed magnitude $m_x$ of stars: $\mu = m_x - M^\prime_x$. By substituting the resulting distance modulus into the standard magnitude equation: $m_x = M_x + A_x + \mu$, we can simulate the magnitude of the stars in the individual passbands. With \teff, \feh, \mg\ and \ebr\ as free parameters, we can model the stellar observed magnitude in each filter. We define,

\begin{equation}
\chi^2 = \dfrac{1}{N-K}\sum^N_{x=1}(\dfrac{(m^{\rm obs}_x-m^{\rm mod}_x)}{\sigma_x})^2, 
\end{equation}
where $m^{\rm obs}_x$ and $m^{\rm mod}_x$ are respectively the observed and simulated magnitudes of the filter $x$, $\sigma_x$ are the photometric errors, $N$ and $K$ is the number of adopted filters and free parameters, respectively. 
If Gaia parallax exists,

\begin{equation}
\chi^2 = \dfrac{1}{N+1-K}(\sum^N_{x=1}(\dfrac{(m^{\rm obs}_x-m^{\rm mod}_x)}{\sigma_x})^2 + (\dfrac{(\varpi_{\rm obs} +\varpi_{\rm zp} -\varpi_{\rm mod})}{\sigma_{\varpi}})^2), 
\end{equation}
where $\varpi_{\rm obs}$ and $\varpi_{\rm mod}$ are respectively the observed and simulated parallax, $\sigma_{\varpi}$ is the parallax error, $\varpi_{\rm zp}$ ($\varpi_{\rm zp} \equiv -0.026$ mas) is the zero point of the observed parallax.

We search for the minimal $\chi^2$ parameters by running a series of \ebr\ values ranging from $-$0.1 to 6.0 in step of 0.02\,mag and all grids in the reference stellar library. We use only the optical filters, i.e. Gaia $G$\gbp\grp\ and SMSS $gri$, to derive the distances of the stars. This procedure yields best-fit values of \teff, \feh, \mg\ and \ebr\ for the individual stars, which will be adopted as the initial parameters of the following MCMC analysis. If the resulted $\chi^2$ are too large ($\chi^2 > 10$), this means that the stellar parameters in our template library do not fit the observed values well. It is possible that the star of concern is not a normal AFGK star, and then this star will not be used further in the subsequent MCMC analysis.

\subsection{Final parameters from the MCMC analysis}

In order to determine the final parameters and their uncertainties for the individual stars, we adopt the MCMC procedure described in \citet{Foreman-Mackey2013}. The initial values for effective temperature \teff,  metallicity \feh, absolute magnitude \mg, and reddening \ebr\ are set according to the values derived from Sect.~\ref{sec:chi}. The maximum likelihood is defined as follows:

\begin{equation}
L = \prod^N_{x=1}\frac{1}{\sqrt{2\pi}\sigma_{x}}\exp\left(\frac{(X^{\rm obs}_x-X^{\rm mod}_x)^2}{2\sigma_x^2}\right)
\end{equation}
where $X^{\rm obs}_x$ and $X^{\rm mod}_x$ are respectively the observed and simulated magnitudes of the filter $x$, or parallax if available.

To run the MCMC analysis, we created 10 walkers and 20 steps chains, discarding the first 5 steps for burn-in purposes. The posterior distributions of the final parameters are determined by the 50th percentile values, and their uncertainties are obtained by computing the 16th and 84th percentile values.
  
\section{Tests of our algorithm}\label{sec:tes}

In this section, we will test the SPar algorithm. We cross-match the SMSS data with the LAMOST, APOGEE and GALAH spectroscopic data, and have obtained a test sample of stars with parameter measurements from the spectroscopic data. The SPar algorithm is then applied to the sample stars to obtain their parameters, which are then compared with those obtained from the spectra. The test sample contains 1,046,722 stars, of which 408,671, 200,902 and 437,149 stars are from the LAMOST, APOGEE and GALAH surveys, respectively.  

\begin{figure*}
\centering
\includegraphics[width=0.5\textwidth,angle=0]{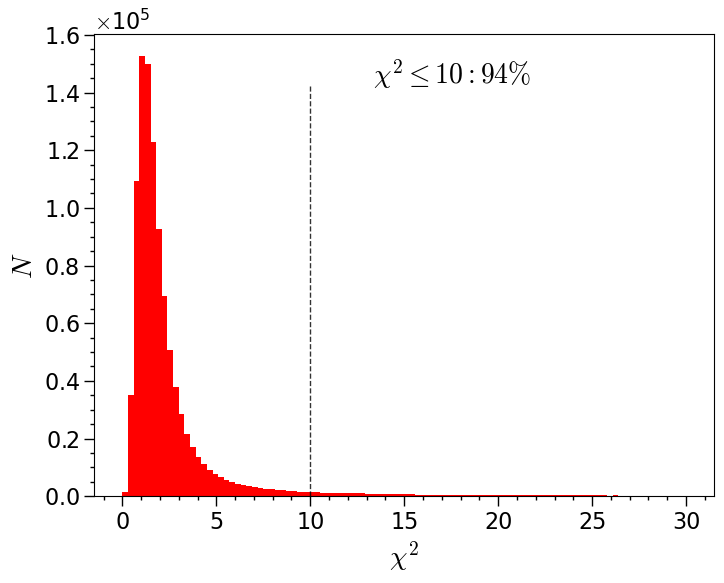}
\caption{The distribution of $\rm \chi^2$. The black dashed line is $\rm \chi^2$=10 line.}  
\label{distr}
\end{figure*}

Fig.~\ref{distr} displays the distribution of $\chi^2$ values for all sources included in the test sample. A typical $\chi^2$ distribution is observed, with a prominent peak at 2 and a long tail. Some stars in the sample exhibit a large $\chi^2$ value, indicating that our empirical models are unable to effectively match their observed data. This may be due to the atmospheric parameters of these stars lying outside the range of our models. To optimize computational efficiency, we have excluded these stars from subsequent MCMC analysis. Consequently, by adopting a conservative $\chi^2$ threshold of less than 10 in our study, we have retained 96\% of LAMOST sources, 85\% of APOGEE sources, and 97\% of GALAH sources.

\begin{figure*}
\centering
\includegraphics[width=1\textwidth,angle=0]{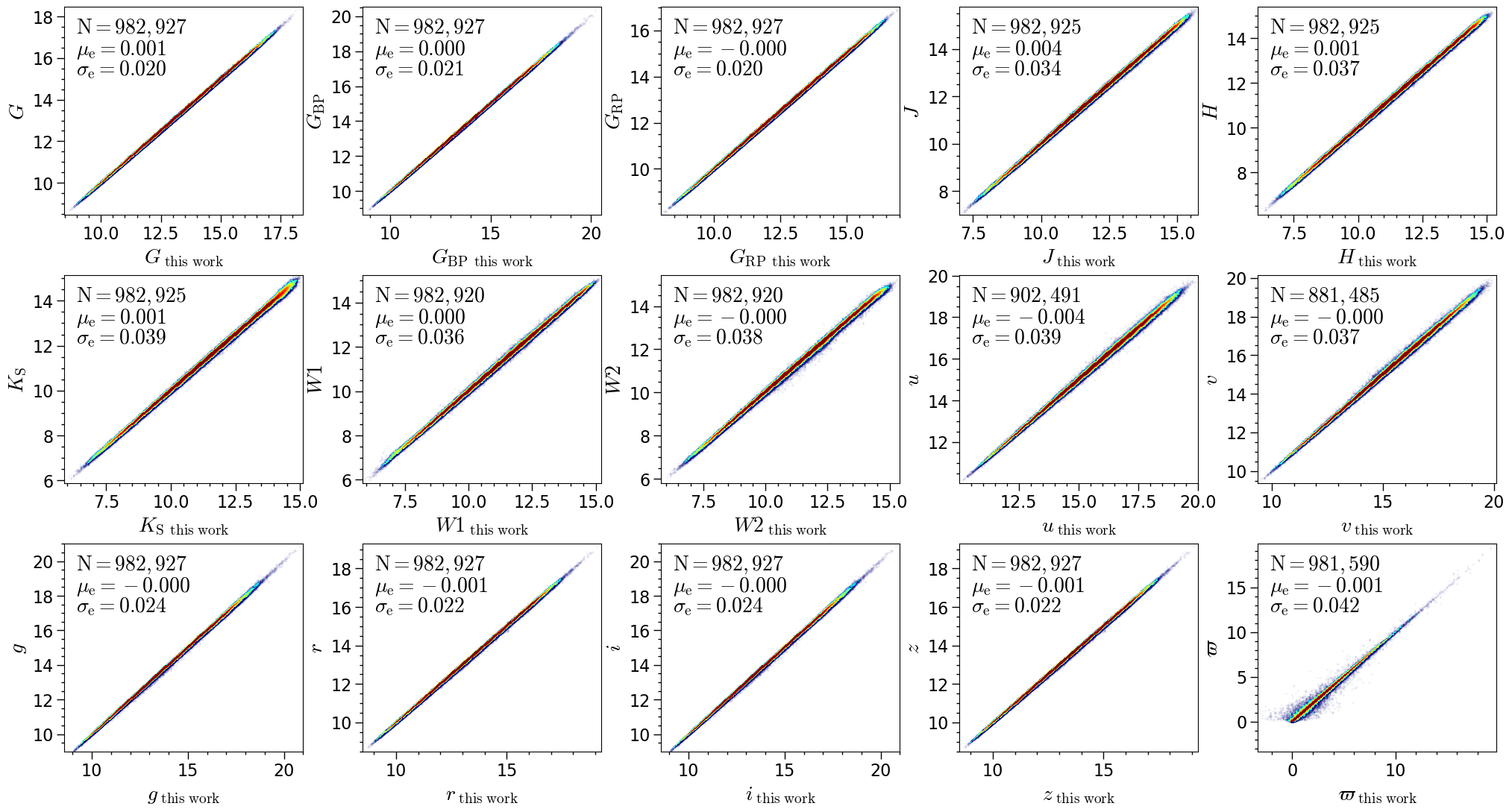}
\caption{Comparison of simulated magnitude and parallax based on the MCMC results with the observed data. N is the total source number, $\mu_{\rm e}$ and $\sigma_{\rm e}$ are the median and standard deviation of differences after 3 sigma elimination.}  
\label{compresd}
\end{figure*}

After conducting the MCMC analysis for the remaining 982,927 stars, we obtained their final parameters. To evaluate the accuracy of our algorithm, we first compared the differences between our predictions and the observations. We simulated the observed magnitudes of the individual stars at various wavelengths and their parallaxes based on the MCMC results, which were compared with the observed data presented in Fig.~\ref{compresd}. Our simulations show good agreement with the observational results. The mean values of the differences are negligible. The dispersions of the differences are about  0.020-0.024 mag for the optical magnitudes ($G$ \gbp \grp $g r i z$), 0.034 - 0.039\,mag for the UV and near-IR magnitudes ($u v J H $\ks$ W1 W2$), and 0.042\,mas for parallax, respectively.

\subsection{Comparison of the resulted parameters}

In this section, we then compare the stellar parameters obtained by SPar to those obtained from the spectroscopic surveys to test the accuracy of Spar. 

\begin{figure*}
\centering
\includegraphics[width=1\textwidth,angle=0]{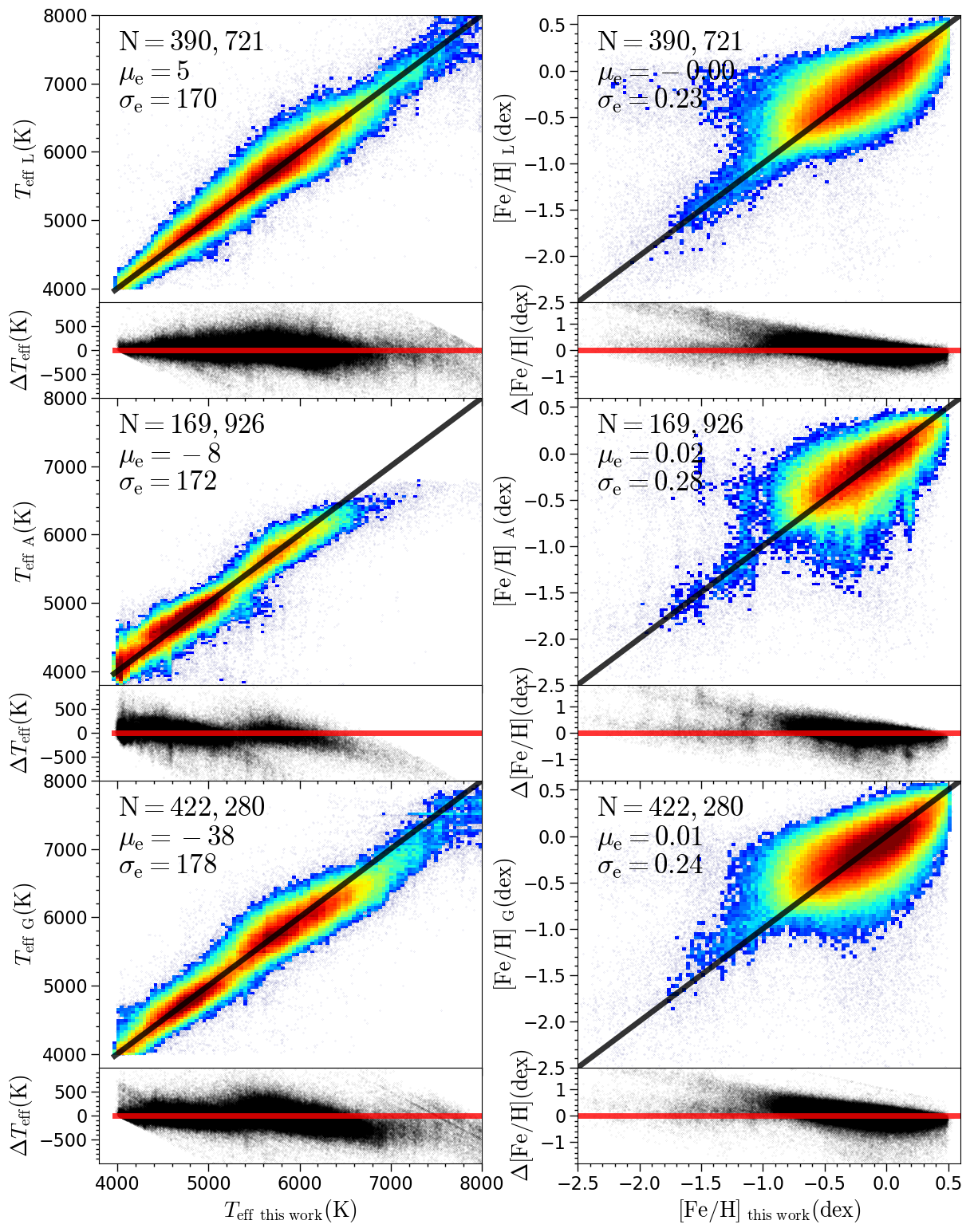}
\caption{Comparison of \teff\ (left) and \feh\ (right) of this work with LAMOST (top), APOGEE (middle) and GALAH (bottom). For every panel, the top half shows the consistency of \teff\ or \feh}, and the bottom half shows the differences change with \teff\ or \feh\ of this work. The black and red line in the top and bottom half panel are the equal and zero lines.
\label{compresa}
\end{figure*}

Fig.~\ref{compresa} shows the comparisons of effective temperature \teff\ and metallicities \feh\ between the current work and the spectroscopic surveys, including LAMOST, APOGEE, and GALAH. The effective temperature from SPar and LAMOST are in good agreement without any obvious trend of change with temperature. The dispersion of the difference between the effective temperatures is only 170\,K. Compared to LAMOST, APOGEE has more low-temperature stars and fewer high-temperature stars. Some of the stars in APOGEE have \teff\ below 4000,K, which is outside the temperature range of our empirical stellar templates. For those low-temperature stars, we would overestimate their temperatures. Regarding the GALAH data, our effective temperatures are systematically higher than the GALAH values for high-temperature stars. The possible reason for this is that the effective temperatures measured by GALAH are systematically higher than those measured by LAMOST for stars of high temperatures, as discussed in \citet{buder2019}.

The sensitivity of the SkyMapper $uv$ filters to stellar metallicities enables accurate measurements of \feh\ based on the SMSS multiband photometric data, as illustrated in Fig.~\ref{compresa}. Our resulting metallicities show good agreement with those obtained from the LAMOST, APOGEE, and GALAH surveys, even for extremely metal-poor stars with \feh\ $\sim$ $-2.5$ dex. However, the dispersion of the differences is relatively high, with values of 0.23, 0.28, and 0.24 dex, respectively, for LAMOST, APOGEE, and GALAH measurements. These values are larger than the dispersion values reported by \citet{Huang2022} (0.05 to 0.15 dex). This difference may be attributed to the fact that \citet{Huang2022} restricted their analysis to stars at high Galactic latitudes, where dust extinction values are small and readily derived from two-dimensional extinction maps. In contrast, many of the stars in our sample are located in the Galactic disk, which is subject to high extinction effects. Therefore, errors arising from dust extinction hinder accurate determination of the intrinsic colors of stars, leading to relatively large uncertainties in the derived metallicities.

\begin{figure*}
\centering
\includegraphics[width=1\textwidth,angle=0]{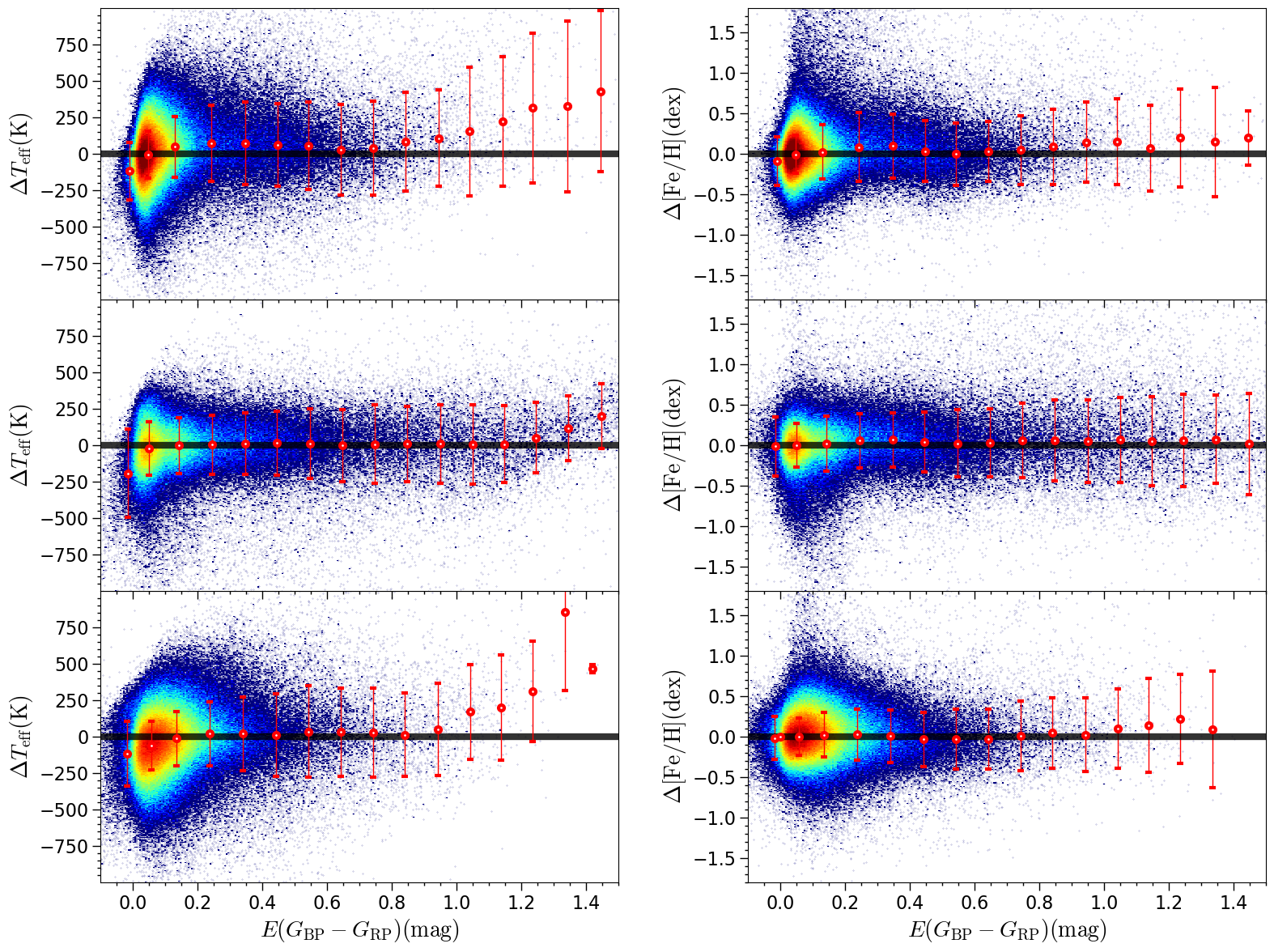}
\caption{The changes of $\Delta$\teff\ (left) and $\Delta$\feh\ (right) with \ebr\ of LAMOST (top), APOGEE (middle) and GALAH (bottom) stars. The red point and error bar is the median and standard deviation in every 0.1 \,mag bin. The black line is the zero line.}  
\label{compresa1}
\end{figure*}

We plotted the differences in effective temperature and metallicities between our work and the spectroscopic surveys against the reddening values of our sample stars in Fig.~\ref{compresa1}. As the reddening values increase, the dispersions of the differences in the two parameters also increase. On average, the mean values of the differences do not vary with the reddening. However, for highly reddened stars, the mean value of the differences in the effective temperature deviates from zero. This may be due to the small number of stars in such regions and the relatively larger errors associated with them.

\begin{figure*}
\centering
\includegraphics[width=1\textwidth,angle=0]{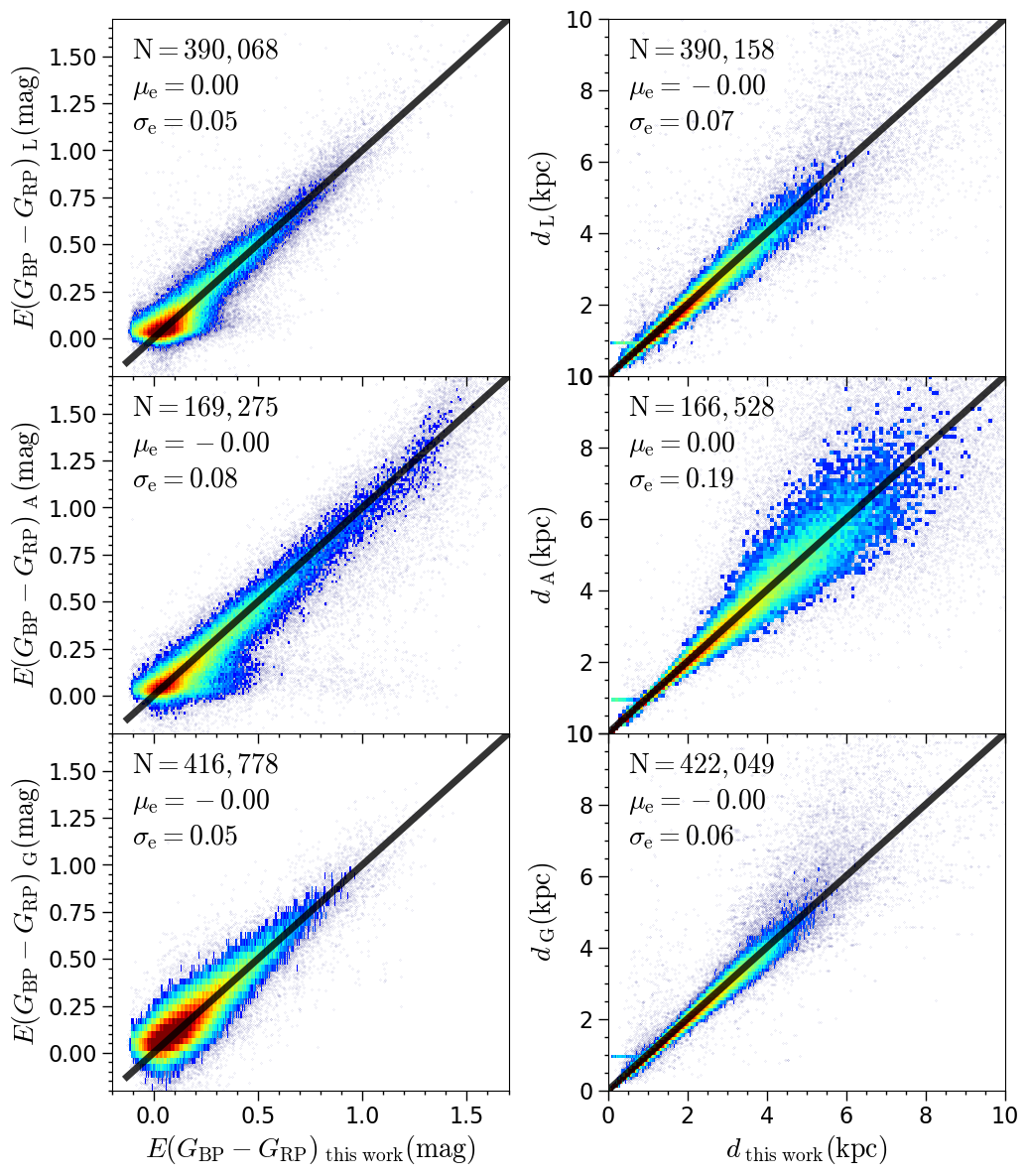}
\caption{The same as Fig.~\ref{compresd}, but for \ebr\ and distance. The black line is the equal line.}  
\label{compresb}
\end{figure*}

In addition to the stellar parameters, reddening \ebr\ and distance are also results from SPar. We compare our resultant \ebr values with those derived from the star-pair method, our derived distances and those from Gaia DR3 parallaxes for all stars in the test sample in Fig.~\ref{compresb}. For \ebr, the consistencies are good. There are no offsets between our results and those from the spectroscopic results. The dispersion values of the differences for the LAMOST and GALAH stars are about 0.05\,mag. While for the APOGEE stars, the dispersion is larger, of about 0.08\,mag. This is partly due to the high proportion of stars with large reddening values in APOGEE, and partly due to the fact that there are some low-temperature stars in APOGEE with temperatures below 4000\,K. As mentioned above, we may overestimate their effective temperatures, which would lead us to overestimate their reddening values at the same time.

Regarding the distance measurements, our results are consistent with the Gaia measurements, with no significant offsets observed. The dispersions of the relative differences for both LAMOST and GALAH stars are only around 7\%, while for APOGEE stars, the value is about 19\%. This is because LAMOST and GALAH stars, being mainly dwarfs that are relatively close to us, have accurate parallax measurements from Gaia, resulting in smaller relative distance dispersions. Conversely, the APOGEE catalogue contains many giant stars that are further away, leading to larger parallax measurement errors and, therefore, a larger dispersion in relative distance measurements.

\begin{figure*}
\centering
\includegraphics[width=1\textwidth,angle=0]{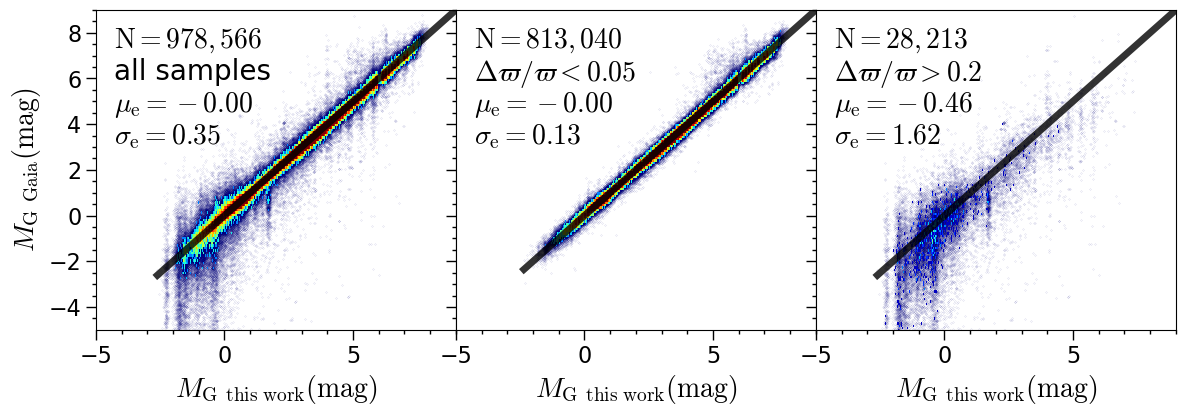}
\caption{The same as Fig.~\ref{compresd}, but for $M_{\rm G}$ of all samples (left), high precision parallax samples (middle), and low precision parallax samples (right). The black line is the equal line.}  
\label{compresc}
\end{figure*}

Finally, we compared the absolute magnitudes in the Gaia $G$ band, $M_G$ values obtained from SPar, with those from the three surveys of all the test stars and found that, in general, the agreement is good (Fig.~\ref{compresc}). The dispersion value of the differences is about 0.35 mag, but the parallax error has a significant impact on the accuracy of the distances, and therefore, it has a great impact on the accuracy of the absolute magnitude we obtain. For stars with small relative parallax errors (less than 5\%), the dispersion value of the $M_G$ difference is only 0.13 mag. However, for stars with relative parallax errors greater than 20\%, these sources exhibit significant dispersion, and the dispersion value of the $M_G$ difference is 1.62 mag.

\subsection{Comparison of results with longer MCMC chains}

\begin{figure*}
\centering
\includegraphics[width=1\textwidth,angle=0]{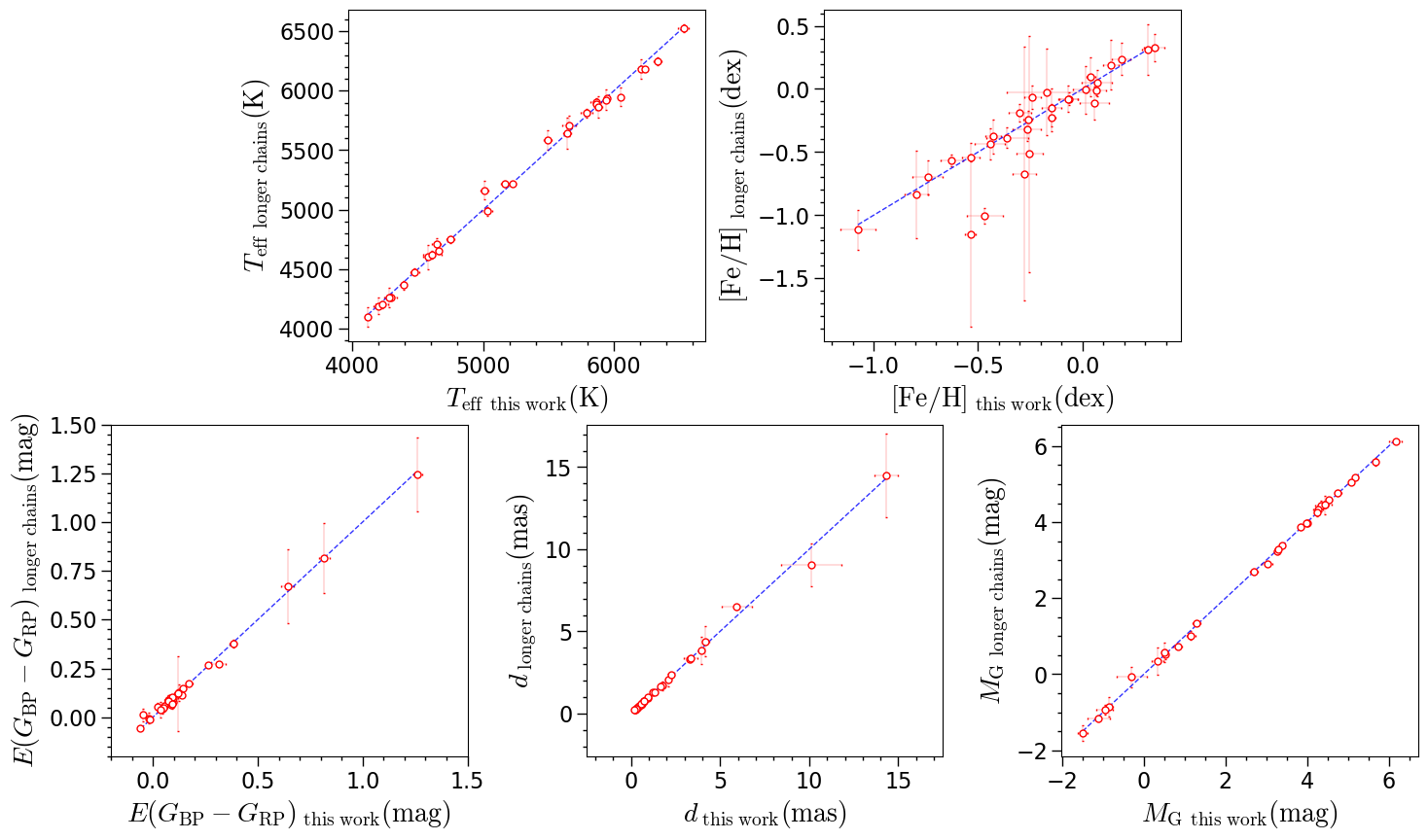}
\caption{Comparison of final parameters (\teff\ (top-left), \feh\ (top-right), $M_{\rm G}$\ (bottom-left), \ebr\ (bottom-middle) and distance (bottom-right)) of this work with parameters from 100 walkers and 1000 steps chains. The blue dashed line is the equal line.}  
\label{compres3}
\end{figure*}

To enable SPar to run on large samples of stars, we employed a smaller number of chains and steps in the final MCMC analysis phase. To evaluate the effect of chain and step numbers on the results, we randomly selected 30 sources (10 sources each from LAMOST, APOGEE, and GALAH) and performed 100 walkers and 1000 step chains for each of these sources. The results obtained from SPar and longer chains and steps are presented in Fig.~\ref{compres3}. Overall, there are good agreements between the obtained parameters. However, four sources showed relatively large deviations in \feh, mainly due to the large uncertainties. Despite this, we believe it is reasonable for SPar to use relatively short chains and steps. Using longer chains and steps can significantly increase the computational time without bringing any significant improvement in the results. 

\section{Summary}\label{sec:sum}

In this work, we have developed a new algorithm called SPar to derive stellar parameters from multi-band photometries, which can be applied to large samples of stars. The algorithm takes advantage of empirical stellar libraries constructed from Gaia, LAMOST, and other photometric surveys. It leverages the minimum $\chi^2$ fit of the stellar SEDs to obtain the initial values for the MCMC analysis, which results in the stellar parameters, including \teff, \feh, \ebr\ and $M_G$, of the individual stars. Our algorithm is tested on the LAMOST, APOGEE, and GALAH stars. The typical dispersion values of the differences between our results and literature values were 170\,K for \teff, 0.23\,dex for \feh, 0.13\,mag for $M_G$ and 0.05\,mag for \ebr.

In the future, our new SPar algorithm will be implemented on large samples of stars obtained from the Mephisto and CSST surveys to derive atmospheric parameters, distances, and extinction values for billions of stars. This will give us crucial insights into the structure, chemistry, and other properties of the Milky Way. 

\section*{Acknowledgements}
We would like to thank the referee for providing us with detailed and constructive feedback that has significantly enhanced the quality of the manuscript. We are grateful to Professor Biwei Jiang for help and discussion. This work is partially supported by the National Key R\&D Program of China No. 2019YFA0405500, National Natural Science Foundation of China 12173034, 11833006, 12203016 and 12173013, Natural Science Foundation of Hebei Province No.~A2022205018, A2021205006, 226Z7604G, and Yunnan University grant No.~C619300A034, and Science Foundation of Hebei Normal University No.~L2022B33. We acknowledge the science research grants from the China Manned Space Project with NO.\,CMS-CSST-2021-A09, CMS-CSST-2021-A08 and CMS-CSST-2021-B03. We are grateful for the support of the Postdoctoral Research Station in Physics at Hebei Normal University.

This research made use of the cross-match service provided by CDS, Strasbourg.

Guoshoujing Telescope (the Large Sky Area Multi-Object Fiber Spectroscopic Telescope LAMOST) is a National Major Scientific Project built by the Chinese Academy of Sciences. Funding for the project has been provided by the National Development and Reform Commission. LAMOST is operated and managed by the National Astronomical Observatories, Chinese Academy of Sciences.

This work presents results from the European Space Agency (ESA) space mission Gaia. Gaia data are being processed by the Gaia Data Processing and Analysis Consortium (DPAC). Funding for the DPAC is provided by national institutions, in particular the institutions participating in the Gaia MultiLateral Agreement (MLA). The Gaia mission website is https://www.cosmos.esa.int/gaia. The Gaia archive website is https://archives.esac.esa.int/gaia.

The national facility capability for SkyMapper has been funded through ARC LIEF grant LE130100104 from the Australian Research Council, awarded to the University of Sydney, the Australian National University, Swinburne University of Technology, the University of Queensland, the University of Western Australia, the University of Melbourne, Curtin University of Technology, Monash University and the Australian Astronomical Observatory. SkyMapper is owned and operated by The Australian National University’s Research School of Astronomy and Astrophysics. The survey data were processed and provided by the SkyMapper Team at ANU. The SkyMapper node of the All-Sky Virtual Observatory (ASVO) is hosted at the National Computational Infrastructure (NCI). Development and support the SkyMapper node of the ASVO has been funded in part by Astronomy Australia Limited (AAL) and the Australian Government through the Commonwealth’s Education Investment Fund (EIF) and National Collaborative Research Infrastructure Strategy (NCRIS), particularly the National eResearch Collaboration Tools and Resources (NeCTAR) and the Australian National Data Service Projects (ANDS).

This publication makes use of data products from the Two Micron All Sky Survey, which is a joint project of the University of Massachusetts and the Infrared Processing and Analysis Center/California Institute of Technology, funded by the National Aeronautics and Space Administration and the National Science Foundation.

This publication makes use of data products from the Widefield Infrared Survey Explorer, which is a joint project of the University of California, Los Angeles, and the Jet Propulsion Laboratory/California Institute of Technology, funded by the National Aeronautics and Space Administration.

\bibliographystyle{aasjournal}
\bibliography{spp}


 \begin{table*}
    \centering
    \begin{tabular}{lrrrrrr}
    \hline
    Band & $C_0$ &$C_1$ & $C_2$ & $C_3$   & $C_4$    & $C_5$ \\
    \hline 
SkyMapper $u$  & $-8.847E-01$ & $1.274E-05$ & $1.175E-07$ & $-4.431E-01$ & $-3.718E-04$ & $-4.343E-01$ \\
SkyMapper $v$  & $-7.932E-01$ & $1.136E-04$ & $8.852E-09$ & $-3.385E-01$ & $-1.186E-04$ & $-2.533E-01$ \\
SkyMapper $g$  & $-1.103E-01$ & $3.665E-05$ & $-1.261E-08$ & $-5.843E-03$ & $2.783E-05$ & $5.486E-03$ \\
SkyMapper $r$  & $4.524E-01$ & $3.872E-05$ & $-1.091E-08$ & $-1.588E-02$ & $2.014E-05$ & $-1.428E-02$ \\
SkyMapper $i$ & $1.008E+00$ & $1.128E-05$ & $-1.602E-08$ & $-1.277E-02$ & $1.162E-05$ & $-1.969E-03$ \\
SkyMapper $z$  & $1.352E+00$ & $-8.916E-06$ & $-8.834E-09$ & $-2.468E-02$ & $1.387E-05$ & $2.657E-02$ \\  
Gaia $G$  & $5.051E-01$ & $-5.638E-05$ & $8.313E-09$ & $6.334E-02$ & $-4.587E-06$ & $-2.592E-03$ \\
2MASS $J$  & $1.787E+00$ & $-3.056E-05$ & $7.046E-09$ & $-4.065E-03$ & $9.801E-06$ & $-2.193E-03$ \\
2MASS $H$  & $2.029E+00$ & $-5.226E-05$ & $8.943E-09$ & $-1.573E-02$ & $2.244E-05$ & $1.496E-02$ \\
2MASS \ks\  & $2.149E+00$ & $-5.576E-05$ & $1.018E-08$ & $-7.487E-03$ & $1.909E-05$ & $9.137E-03$ \\
WISE $W1$  & $2.235E+00$ & $-6.046E-05$ & $1.438E-08$ & $1.043E-02$ & $1.891E-05$ & $-3.675E-03$ \\
WISE $W2$  & $2.276E+00$ & $-7.125E-05$ & $1.153E-08$ & $-2.593E-03$ & $3.199E-05$ & $9.640E-03$ \\
    \hline
    \end{tabular}
    \caption{Temperature- and reddening-dependent reddening coefficients of the individual passbands. The function form is: $R = E(G_{BP}-x)/E(G_{\rm BP}-G_{\rm RP}) = C_{0}+C_{1}x+C_{2}x^2+C_{3}y+C_{4}xy+C_{5}y^2$, where $x = T_{\rm eff} - 6000$ and $y = E(G_{\rm BP}-G_{\rm RP}) - 0.5$.}
    \label{tab1}
\end{table*}

\end{CJK*}
\end{document}